\documentclass[aps,twocolumn]{revtex4}
\usepackage{amsmath,amssymb,graphicx}
\usepackage{algorithmic}
\usepackage{enumerate}

\begin{document}

%\date{\today}
%\maketitle

\title{\textbf{Transitions From Order to Disorder in Multi-Dark and 
Multi-Dark-Bright Soliton Atomic Clouds}}

\author{Wenlong Wang}
\email{wenlong@physics.umass.edu}
\affiliation{Department of Physics, University of Massachusetts,
Amherst, Massachusetts 01003 USA}

\author{P.G. Kevrekidis}
\email{kevrekid@math.umass.edu}
\affiliation{Department of Mathematics and Statistics, University of Massachusetts,
Amherst, Massachusetts 01003-4515 USA}

\affiliation{Center for Nonlinear Studies and Theoretical Division, Los Alamos
National Laboratory, Los Alamos, NM 87544}

\begin{abstract}
We have performed a systematic study quantifying the variation of 
solitary wave behavior from that of an ordered cloud resembling
a ``crystalline'' configuration to that of a disordered state 
that can be characterized as a soliton ``gas''. As our illustrative examples,
we use both one-component, as well as two-component, one
dimensional atomic gases
very close to zero temperature, where in the presence of repulsive
inter-atomic interactions and of a parabolic trap, a cloud, respectively
of dark (dark-bright) solitons can form in the one- (two-) component system.
We corroborate our findings through three distinct types of approaches,
namely a Gross-Pitaevskii type of partial differential equation,
particle-based ordinary differential equations describing the 
soliton dynamical system and Monte-Carlo simulations for the particle
system. We define an ``empirical''
order parameter to characterize the order of the soliton lattices and study how this changes as a function of the strength 
of the ``thermally'' (i.e., kinetically) induced perturbations. As may be
anticipated by the one-dimensional nature of our system, the transition
from order to disorder is gradual without, apparently, a genuine 
phase transition ensuing in the intermediate regime. 
\end{abstract}

\maketitle

\section{Introduction}

The theme of nonlinear waves, and their dynamics and interactions
has amply blossomed over the past two decades in the realm of
atomic Bose-Einstein condensates (BECs)~\cite{book1,book2}. This is 
because BECs enable the experimental realization of both 
focusing and defocusing nonlinear Schr{\"o}dinger type models
in the form of the Gross-Pitaevskii equation at near-zero
temperature for atomic gases with, respectively, attractive or
repulsive inter-atomic interactions~\cite{emergent}. It is for
that reason that a diverse array of structures encompassing,
but not limited to, bright solitary waves~\cite{expb1,expb2,expb3}, gap matter
waves~\cite{gap} dark solitons~\cite{emergent,djf},
vortices~\cite{emergent,fetter1,fetter2}, as well as solitonic
vortices and vortex rings~\cite{komineas_rev} have been explored
in this context. 

Dark solitons in one-component repulsively self-interacting BECs represent
one of the most intensely studied coherent structures.
Early experiments in this context~\cite{han1,nist,dutton,han2} were,
at least in part, limited by dynamical
instabilities affecting the lifetime of the states 
in higher dimensional settings, as well as by the role
of thermal fluctuations at temperatures closer
to the transition temperature. 
More recent experiments, however, have been able to provide
a significantly increased control over the formation and
dynamical evolution of such 
states~\cite{engels,Becker:Nature:2008,hambcol,kip,andreas,jeffs}.
By combining sufficiently low temperatures 
and closer to one-dimensional regimes, a number of these
more recent experimental efforts have provided 
clear imprints of oscillating and interacting
dark solitons, in good agreement with theoretical predictions.

One of the additional remarkable features of the BEC realm is
that it controllably enables the consideration not only of 
the single-component system, but also of multi-component
ones, e.g., consisting of different hyperfine states of
the same atomic gas such as $^{87}$Rb~\cite{book1,book2,emergent}.
In the latter setting, one of the particularly relevant
dynamical structures  experimentally realized (in the repulsive interatomic 
interaction regime) are the
so-called dark-bright (DB) solitary waves. These were initially produced
in optical settings~\cite{seg1,seg2,seg3}, yet subsequently gained
considerable momentum in the BEC realm due, again, to the wide
and diverse as well as robust array of epxeriments that produced
them~\cite{sengdb,peterprl,peter1,peter2,peterpra,peter3}.
The remarkable feature about this structure is the fact that 
while bright solitons cannot
exist ``on their own'' in the repulsive interatomic interaction case
within DB solitary waves, the dark solitary structures 
play the role of an effective 
potential that enables the bound state trapping of the bright component. 
As a result, robust DB states have been observed
to oscillate in a parabolic trap ~\cite{sengdb,peter1}, to be 
spontaneously produced by  counterflow experiments~\cite{peterprl}, and 
to form bound states~\cite{peter2}.
It is also worthwhile to note that SU$(2)$-rotated
siblings of DBs have also been experimentally
observed in the form of beating dark-dark solitons~\cite{peterpra,peter3}.

While the dynamics of few solitary waves is most typically studied
in the above works (and their dynamical robustness, where appropriate,
is established) far fewer studies have concerned themselves with
the properties of large cohorts of such waves and their potential
states including e.g. a crystalline equilibrium state or a 
disordered highly interacting (and perhaps chaotic) state.
Nevertheless, the topic of transitions from a soliton ``crystal''
to a soliton ``gas'' is a fairly old one; see e.g. for a 15-year
old discussion the work of~\cite{mitchke}. Additionally, it is one
that has been meeting with renewed interest not only in
single component settings, but also in multi-component ones;
see e.g. the recent discussion about different phases (including
a topological Wigner crystal) of half-solitons (in our setting,
DB ones) of~\cite{tercas}. On the other hand, a considerable
attention has been paid to far from equilibrium phenomena
such as turbulent dynamics (i.e., ``soliton turbulence'')
and their relaxation~\cite{el1,el2,zakharov,gasenzer}. 

Our aim in the present work is to revisit the experimentally
tractable setting of one- and two-component atomic BECs and
consider large arrays of coherent structures in the form of
dark solitons (see e.g. for a recent example of a relevant
discussion~\cite{DS1}), and dark-bright solitons (see e.g.
for a recent example~\cite{tsitoura}). For these arrays,
we intend to describe ``transitions'' between ordered, 
crystalline-type states to disordered, gaseous-type
states. Notice that we do not identify phase transitions 
by means of our diagnostics, a feature that appears to be
fairly plausible given the  one-dimensional nature of our
system. We devise a suitable order parameter, measuring the 
deviation of the different states from their respective
equilibria and explore these states as a function of a kinetically
defined temperature. Our indication about the absence of a genuine
phase transition arises in the form of a smooth, continuous
dependence of the order parameter on our ``kinetic temperature''.
Nevertheless, we cannot exclude the possibility that our choice
of order parameter may be the one that precludes the identification  
of a phase transition. Nevertheless, we believe that the identification
of such dynamical states (resembling ``solitonic crystals'' and
``solitonic gases'') will be valuable in prompting the
further development of both theoretical
and experimental tools to explore them.

Our presentation is structured as follows. In section II, we introduce
the single-component setting of dark solitons and explain our 
three-fold computational approach: (a) based on the partial differential
equation (PDE) of the GPE type; (b) based on the ordinary differential
equation (ODE) describing the solitary waves as particles and
finally (c) based on a population annealing Monte Carlo (PAMC) approach for the
particle system (consisting of the solitons). In section III, we
present corresponding information about the dark-bright states
and two-component BECs . In section IV, we collect our numerical results
about the order-disorder transition as our kinetic temperature is
varied in all three of the above approaches, for each of the two
systems. Finally, in section V, we summarize our findings and
present a number of directions for future study.

\section{One component dark solitons}
\label{DS}
\subsection{Models and the particle picture}

Our examination of the dark soliton system will take place 
in the large density limit, where the equilibrium positions are 
known and can be identified for an arbitrary number of coherent
structures~\cite{DS1}. We model the dark solitons using the repulsive 1d 
GPE equation with a harmonic potential. The GPE equation can be
written as (assuming for computational simplicity a trap frequency of 
unity, although our considerations are fully generalizable to 
the case of arbitrary trap strength; for a discussion of the
reduction to the 1d model see, e.g.,~\cite{emergent})
\begin{equation}
i \psi_t=-\frac{1}{2} \psi_{xx}+\frac{1}{2} x^2 \psi +| \psi |^2 \psi-\mu \psi.
\label{GPE}
\end{equation}
Here, $\psi \in \mathbb{C}$ is a complex field defined on $(x,t)\in (\mathbb{R},\mathbb{R})$ and $\mu$ is the chemical potential related to the total number of particles in the BEC.

The static properties and the low lying dynamical normal mode frequencies were 
explored in detail in the particle picture in the large density limit in \cite{DS1}. We will briefly summarize some of the key results for our 
subsequent discussion herein. A scaling transformation of Equation~(\ref{GPE}) 
can be selected to yield the semi-classical form of the nonlinear
Schr{\"o}dinger model:
\begin{equation}
\psi=\sqrt{\mu}u,\ x=\sqrt{2\mu}\xi,\ t=2\tau,\ \epsilon=\frac{1}{2\mu}.
\end{equation}
Equation~(\ref{GPE}) then becomes
\begin{equation}
i\epsilon u_{\tau}+\epsilon^2 u_{\xi\xi}+(1-{\xi}^2-|u|^2)u=0.
\label{GPE1}
\end{equation}

In the limit $\mu \rightarrow \infty$ or equivalently $\epsilon \rightarrow 0$, Equation~(\ref{GPE1}) has a limiting static solution
\begin{equation}
\eta(\xi)=(1-{\xi}^2)^{1/2},
\end{equation}
with $\xi  \in [-1,1]$.
We will call the former space the real space and the 
latter space the scaled space in this work, for reference. 

It is an interesting fact that particle-like excitations can be ``baptized'' 
on top of the BEC background in the dark solitonic form
\begin{equation}
v(\xi,\tau)=A \tanh(\epsilon^{-1} B(\xi-a))+ib,
\label{solitonform}
\end{equation}
where $A=\sqrt{1-b^2},\ B=\sqrt{\frac{(1-a^2)(1-b^2)}{2}}\ \rm{and}\ a \in (-1,1), b \in (-1,1).$
As is well-known~\cite{djf}, $a$ represents the position of the dark soliton
while $b$ corresponds to its velocity. 
The number of dark solitons that can be meaningfully fit within the
domain is only limited by the number of healing lengths (the characteristic
size of the soliton~\cite{emergent}) that can be placed within the radius
of the static solution; yet, by suitable tunning of the trap and of the
chemical potential, this number can be made arbitrarily large. 
Hence, in general, one can grow $s$ dark solitons by multiplying $s$ 
equations in the form of Equation~(\ref{solitonform}), 
with different initial positions and speeds, where $s$ is a positive integer. Then, a general initial state of a system with $s$ dark solitons can be 
written as:
\begin{equation}
u(\xi)=\eta(\xi) \prod_{j=1}^s v_j(\xi,a_j,b_j).
\end{equation}

In \cite{DS1} the equilibrium positions of the dark soliton were identified 
and the effective interactions between them when treated as classical particles can be described using a Toda potential in the form:
\begin{equation}
U(\xi_1,\xi_2)=8e^{-\sqrt{2} \epsilon^{-1} |\xi_1-\xi_2|}.
\end{equation}
We derived by the scaling transformation how the kinetic energy $E_k$ of a dark soliton in real space can be represented in terms of $b$ in the scaled space and the form of the potential energies between the dark solitons $U$ and that between the dark solitons and the trapping potential $V$ in real space. The trapping potential has an effective frequency $\omega_{os}=\frac{1}{\sqrt{2}}$ for dark solitons in the real space (as is well-known~\cite{djf}, this is scaled
by the trap frequency) in the large chemical potential limit
that we are presently considering. The results are summarized as follows
\begin{eqnarray}
\label{trans1}
E_k &=&\frac{\mu}{2} b^2\\
\label{trans2}
U(x_1,x_2)&=&4\mu e^{-2\sqrt{\mu}|x_1-x_2|}\\
\label{trans3}
V(x)&=&\frac{1}{2}\omega_{os}^2 x^2.
\end{eqnarray}

In the rest of this section, we will define the order parameter $m$ and talk about the procedures of the PDE, ODE and the PAMC simulations in detail that will
lead to the characterization of our order-disorder transitions.

\subsection{The order parameter $m$}
Let the positions of each solitary wave
particle be denoted by $\{x_i\}$. Then, order vs. disorder is reflected 
in the \textit{relative positions} between the coherent structures. This 
motivates us to define an order parameter $m$ using  the
relative position normalized by 
the minimum reciprocal wavevector, which is $k_{\rm{min}}=\frac{2\pi}{a_0}$, where $a_0$ is the lattice constant. In particular, our
selection of $m$ is defined as
\begin{eqnarray}
m &=&\frac{\sum\limits_{i=1}^{N-1} \cos(k_{\rm{min}} (x_{i+1}-x_i))}{N-1} \\
&=&\frac{\sum\limits_{i=1}^{N-1} \cos(\frac{2 \pi}{a_0} (x_{i+1}-x_i))}{N-1}
\end{eqnarray}

From the definition of $m$, we can see that $m$ should be expected to
go to zero in the disordered regime, given the fluctuations from
the equilibrium distance and instead to tend to one in the 
ordered regime. This constitutes our motivation for the empirical
selection of
this particular order parameter. As we will see, this will be a useful
tool towards identifying transitions from ordered to disordered regimes
(although this transition will be found to be smooth rather than 
one directly involving or indicating a phase transformation). Nevertheless,
while this is a first step towards quantifying these types of transitions,
it also poses the broader question of identifying suitable diagnostics
for characterizing the phenomenology of these effective particle-wave
entities embedded within an extended infinite-dimensional dynamical 
system.

\subsection{The PDE and ODE simulation}
\label{peak}
We start by summarizing the PDE simulation parameters for dark solitons:
$$\mu=50,\ N=30,\ dx=0.05,\ dt=0.002,\ t=40,$$
where the quantities are chemical potential (chosen to be large to ensure that
a particle description is suitable~\cite{DS1}), number of dark solitons
(also chosen to be reasonable so that averaged quantities can be suitably
defined), spatial and temporal discretization size (chosen for our PDE simulations
to be insensitive to their slight variations) and total simulation 
time, respectively, for the reported results. Our ODE simulation parameters are the same except $dt=0.01$.
We study the time evolution of the state by using the classical RK4 method in 
time and a 2nd-order centered difference discretization scheme in space. 
The dark solitons are first initialized at their equilibrium positions, but with random velocities. For the PDE simulation, we first initialize the state in the scaled space and then transform it to real space. For the ODE simulation, we use the potential energies given in Equation~(\ref{trans2}) and (\ref{trans3}) to
perform the time evolution and compute the kinetic energy from the equation of motion. Since the velocities were initialized with random speeds, we studied many realizations with different initial velocities; each realization will be hereafter termed 
a sample.
The average initial kinetic energy per particle $e_k$ is calculated as
\begin{equation}
e_k=\frac{\mu \sum\limits_{i=1}^{N} b_i^2}{2N}.
\end{equation}
For each sample, we record $e_k$ and measure the state and the order parameter $m$ over each time period 0.1. For the ODE simulation, we do the same but record states at each time step. Note that the $dt$ for PDE is much smaller than that of ODE and it is much more expensive to save a PDE state than an ODE state. Since there are fluctuations of the distribution of speeds for the same $e_k$, we therefore look for statistical relations between $e_k$ and $m$.

\begin{figure}
\begin{center}
\includegraphics[scale=0.46]{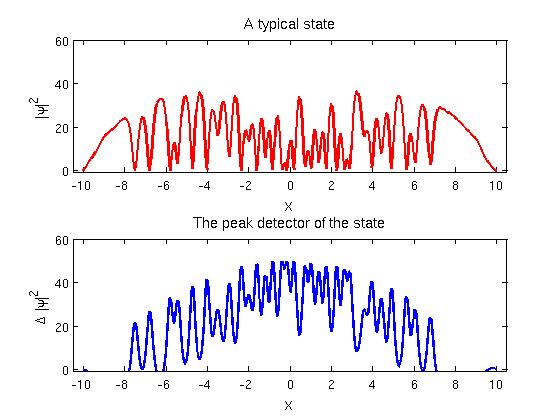}
\caption{(Color Online) A typical state $|\psi|^2$ and dark soliton location detection function $\Delta|\psi|^2$ of the BEC with 30 dark solitons during the system's time evolution.  The upper panel is the state $|\psi|^2$ and the lower panel is the function $\Delta|\psi|^2$ represented by means of a cubic spline interpolation.}
\label{DSstate}
\end{center}
\end{figure}

The positions of the dark solitons for the PDE simulation are extracted from a dark soliton location detection function $\Delta|\psi|^2=|\psi_{\rm{background}}|^2-|\psi|^2$, where $\psi_{\rm{background}}$ is the ground state. We compute this function and do a cubic spline interpolation on a spatial grid of size $0.01$. Then the positions of the dark solitons correspond to the maximum values of the function $\Delta|\psi|^2$ instead of the minimum values of the function $|\psi|^2$, which is more difficult to deal with since $\psi \rightarrow 0$ at the boundaries too. To prevent small  peaks stemming from boundary noise to be recorded, we used an extra condition to require $\Delta|\psi|^2>10$. A typical state and the function $\Delta|\psi|^2$ is given in Figure~\ref{DSstate}.

\subsection{Population annealing Monte Carlo}
A state of the system is a list of $\{x_i\}$.  These wave-particles 
interact with each other and with the trapping potential. Their dynamics
corresponds to an effective one-dimensional lattice, which, in turn,
enables us to utilize statistical mechanics techniques.
We are going to use a Monte Carlo (MC) method to simulate the system. To work in the same state space, we have therefore integrated out the kinetic energy 
of the system. The energy function of the system for the MC simulation thus comes only from  the potential energy. Therefore the temperature should be of the same scale as the potential energy per particle. Then the transition in temperature should be analogous to the transition in kinetic energy.

We initialize the system from the equilibrium positions, although this is not 
necessary. We would like to mention here that it is not necessary to know the lattice constant from the mathematical set up to perform the MC simulation. 
Similarly to the ODE simulation, all we need is the form of 
the inter-particle potentials. The lattice constant can be estimated from the MC method by looking at states at the lowest temperature. 
In this way, we obtained the value of 0.3872 from the MC method 
for the lattice constant compared with the value of 0.3829 of the 
steady state computation of the ODE problem. The two results agree very well. We use our MC lattice constant to compute $m$ (so as 
to make the MC simulation more self-consistent/self-contained). 
The state space of the system is continuous, so it is important to know how to update the state of the system. After some tests with a simple harmonic oscillator, we find that the following MC update method is efficient in the present
setting.
\begin{itemize}
\item Propose a move of random length with a random direction. \\
We can use a random number in the interval $[-h,h]$, where h is a step length and $h>0$, to propose a move to a particle. If the number is positive, then
the move is to the right, while if it is negative, the move is to the left.
\item Update the state using the Metropolis algorithm. \\
We use the Metropolis algorithm in our simulation, i.e., we 
compute the energy change of $dE$ and accept the move with probability $p=\min[1,e^{-\beta dE}]$. Here $\beta$ is the inverse temperature and is related to temperature $T$ as $\beta=\frac{1}{T}$. {In addition, a move is practically rejected if it proposes a swap between two particles since the interaction between them
is strongest when they are on top of each other even for the highest 
temperatures in the simulation. This is not necessary for the 
thermodynamics of the system, but can nevertheless better reflect the dynamics of the system and also simplify the relevant implementation.}
\end{itemize}
In this work, we use $h=1$. We propose the elementary move to all particles sequentially. A Monte Carlo sweep is an update of the elementary moves for all $N$ particles at once. We will use Monte Carlo sweeps to quantify the amount of work we did in our simulations.

Having introduced how to perform a Monte Carlo sweep at an arbitrary temperature, we now introduce the population annealing Monte Carlo algorithm. This algorithm was introduced in \cite{F}. It is an example of sequential Monte Carlo \cite{A}, in contrast to the Markov chain Monte Carlo (MCMC). It is related to simulated annealing, but does annealing with resampling to stay in thermal equilibrium. PAMC has recently been developed and shown to be an efficient algorithm for systems with complicated energy landscapes like spin glasses \cite{TBC,pamc}. 
In this work, we find that PAMC is also efficient for the classical Toda lattice model. The main advantage of PAMC over the simple MC is that PAMC can more accurately
maintain thermal equilibrium even for systems with complicated energy landscapes and thermodynamic quantities at multiple temperatures, often a few hundred, can be obtained in a single run. Also, PAMC can be readily done with parallel computing. In fact, in our work, we used OpenMP for the MC simulation
implementation.

The PAMC algorithm works as follows:
\begin{enumerate}
\item Initialization: Start with $R_0$ independent replicas. Choose $N_T$ temperatures. The highest temperature for spin glasses is often chosen as $\beta=0$. Here, we start from a finite but high temperature. In this way, we can initialize the particles at the equilibrium positions and do some MC sweeps to start the population at thermal equilibrium. In our simulation, we start at $T=5$ with 40 sweeps for each replica and go down to $T=0.1$, where the particles are ordered. PAMC works by decreasing the temperature slowly from a high temperature to a low temperature following an annealing schedule. Here, we use a schedule of even spacing in $\beta$.
\item Resampling: Suppose we are lowering the temperature from $\beta$ to $\beta^{'}$, where $\beta^{'}>\beta$. The re-weighting factor of replica $i$ with energy $E_i$ is proportional to $e^{-(\beta^{'}-\beta)E_i}$ and the expected number of copies of replica $i$ is given by
\begin{equation}
\rho_i(\beta,\beta^{'})=\dfrac{e^{-(\beta^{'}-\beta)E_i}}{Q(\beta,\beta^{'})},
\end{equation}
where Q is just the sum of all the re-weighting factors, divided by $R_0$ to make the sum of $\rho_i$ equal to $R_0$.
\begin{equation}
Q(\beta,\beta^{'})=\frac{\sum\limits_{i=1}^R e^{-(\beta^{'}-\beta)E_i}}{R_0}.
\end{equation}
The number of replicas can be fixed to a constant by using the multinomial distribution \cite {B} or the residual resampling method \cite{math}. We can also allow the population size to fluctuate. For example, we can choose the number of copies for replica $i$ from a Poisson distribution \cite{B} or the nearest integer distribution. Here we use the nearest integer resampling method, which has the smallest variance. Let the integer part of $\rho_i$ be $k_i$. The number of copies $n_i$ of replica $i$ is either $k_i+1$ with probability $\rho_i-k_i$ or $k_i$ with probability $1-(\rho_i-k_i)$. Note that the expectation value of $n_i$ is $\rho_i$.

\item MCMC sweeps: Because the new population is now more correlated, since some of the replicas are the same due to duplications and also for the purpose of ergodicity to fully explore the phase space, since the population size is finite, we do $N_S$ MCMC sweeps to all replicas using the Metropolis algorithm after the resampling step.
\item Repeat step 2 and step 3 $N_T-1$ times to go from the highest temperature to the lowest temperature.
\end{enumerate}
The parameters of the MC simulation of the 30 dark solitons are: $R_0=5\times10^4,\ N_T=301$ and $N_S=10$. Having passed the equilibrium criteria of \cite{TBC}, we believe that we have equilibrated the system at all temperatures.

\section{Dark-bright solitons}
\label{DBS}
\subsection{The coupled GPE model and the particle picture}

As indicated also in the Introduction, dark-bright soliton (DBS) states are 
interesting non-linear structures on top of the BEC background for the one dimensional two-component BECs. 
As such, these states have also undergone intense theoretical 
investigation; see e.g.~\cite{buschanglin,DDB,kanna,rajendran,val,berloff,VB,Alvarez,vaspra,vasnjp,fot} for a number of relevant studies. 
%The model is well described by the mean filed coupled Gross-Pitaevskii equation (GPE). There are two condensate fields.
The prototypical one-dimensional model where such states
can be found to arise is the coupled GPE~\cite{peter2} of the form:
\begin{equation}
i{\psi_j}_t=-\frac{1}{2}{\psi_j}_{xx}+\frac{1}{2}\omega^2x^2{\psi_j}+(|\psi_1|^2+|\psi_2|^2) \psi_j-\mu_j \psi_j,\ j=1,2
\label{BEC}
\end{equation}
where $\psi_j\in \mathbb{C}$ is a complex field of the component $j$ defined on $(x,t)\in (\mathbb{R},\mathbb{R})$ and $\mu_j$ is the chemical potential related to the total number of particles of the component $j$ in the BEC, while 
$\omega$ is the frequency of the trapping potential.
Here, we have also assumed a scattering length setting akin to that
the case of $^{87}$Rb where the near equality of self- and cross-
scattering lengths makes it a reasonable first order approximation
to assume that all the nonlinear prefactors are equal.

A similar transformation can be done to the coupled GPE equation,
as per the discussion of~\cite{peter2}. Here, on top of the 
(inverted parabola) ground
state,
we can ``baptize'' DB solitons of the form: 
\begin{eqnarray}
u_1(\xi,\tau) &=&\cos\phi\tanh\{D[\xi-\xi_0(\tau)]\}+i\sin\phi \\
u_2(\xi,\tau) &=&\eta {\rm{sech}}\{D[\xi-\xi_0(\tau)]\}e^{ik\xi+i\theta(\tau)}.
\end{eqnarray}
In the unperturbed (e.g., by the parabolic trap) problem
the parameters satisfy the following relations
\begin{eqnarray}
D^2 &=&\cos\phi^2-\eta^2 \\
\dot{\xi_0} &=& k = D\tan\phi \\
%k&=&D\tan\phi \\
\theta(\tau)&=&\frac{1}{2}(D^2-k^2)\tau+(\tilde{\mu}-1)\tau,
\end{eqnarray}
where $\tilde{\mu}=\mu_2/\mu_1$.
As before, far from the linear limit, in the so-called Thomas-Fermi regime, 
we can multiply particle-like dark soliton states to the ground state to get the first component (approximate) wavefunction. In the second component, 
we correspondingly add bright soliton states (located at the same spot
as the dark solitons), possibly with a phase difference $e^{i\Delta\theta}$ 
between adjacent waves. A general state with $s$ dark-bright solitons located at $\{a_i\}$ with dark soliton phase angles $\{\phi_i\}$ and bright soliton amplitudes $\{\eta_i\}$ can therefore be written in the following form
\begin{eqnarray}
\psi_1 &=&\sqrt{\mu_1-V}\prod_{j=1}^s \{ \cos\phi_j\tanh\{D_j[\xi-a_j(\tau)]\} \nonumber \\
&+&i\sin\phi_j \}
\end{eqnarray}
\begin{eqnarray}
\psi_2 &=&\sum_{j=1}^s \eta_j {\rm{sech}}\{D_j[\xi-a_j(\tau)]\}e^{ik_j\xi+i\theta_j(\tau)} e^{i\Delta\theta j}.
\end{eqnarray}
If $\Delta\theta=0$ between adjacent waves,  the bright solitons are in phase, while if $\Delta\theta=\pi$ between them, we say the bright solitons are out of phase.
The interaction energy between a pair of \textit{identical} static dark-bright solitons has been recently derived in~\cite{peter2}. 
More specifically, it was indentified in that work that
$U_{DBS}=U_{DD}+2U_{DB}+U_{BB}$, where the three terms stand for 
dark-dark soliton same-component interaction, dark-bright soliton inter-component 
interaction and bright-bright soliton same-component interaction respectively. 
Here, we summarize the kinetic energy for the PDE simulation and the potential energy for the ODE and MC simulations in real space for the dark-bright solitons~\cite{peter2}.

\begin{eqnarray}
\label{EkDBS}
E_k &=&\frac{1}{2}\mu_1 k^2 \\
\label{VDBS}
V(x)&=&\frac{1}{2}\omega_{os}^2x^2 \\
\label{UDD}
U_{DD}&=&\frac{1}{\chi_0}\left(\frac{272-176D_0^2}{3D_0}+32(D_0^2-1)(r+\frac{1}{2D_0})\right) \nonumber \\
&\times& e^{-2D_0r} \\
\label{UBB}
U_{BB}&=&\frac{\chi}{\chi_0}\left(-6D_0-2\chi+2D_0^2(r+\frac{1}{D_0})\right) \nonumber \\
&\times& D_0\cos(\Delta\theta)e^{-D_0r}+\frac{\chi^2}{\chi_0}(1+2\cos^2\Delta\theta) \nonumber \\
&\times&\left(3-2D_0(r+\frac{1}{2D_0})\right)D_0e^{-2D_0r} \\
\label{UDB}
U_{DB}&=&\frac{\chi}{\chi_0}8\cos(\Delta\theta)e^{-D_0r} \nonumber \\
&+&\frac{\chi}{\chi_0}\left(-\frac{104}{3}+16D_0(r+\frac{1}{2D_0})\right)e^{-2D_0r},
\end{eqnarray}
where $\omega_{os}^2=\omega^2\left(\frac{1}{2}-\frac{\chi}{\chi_0}\right)$. $D_0$ is the value of $D$ for the static dark-bright solitons, $\chi=\frac{2\eta^2}{D_0}$ and $\chi_0=8\sqrt{1+\left(\frac{\chi}{4}\right)^2}$. $r$ is the distance between the two adjacent dark-bright solitons. We can also define the average initial kinetic energy per DBS, $e_k$. It is clear that the interaction of the DBSs is much more complicated than that of the single-component dark solitons. The interaction depends on the amplitudes of the bright solitons via $\chi$.
Therefore, to perform the relevant 
simulations using the particle picture, we need some input of the parameters of $\eta$ of each bright soliton. We extract this information from the numerical stationary DBS state. We can optimize the unknown parameters of the particle state by minimizing the norm of the difference of the particle state and the numerical stationary state. Therefore, we will first talk about an effective procedure to obtain multi-DBS stationary states in the next section.

\subsection{Identification and continuation of stationary DBS states}
\label{DBSss}
\begin{figure}
\begin{center}
\includegraphics[scale=0.46]{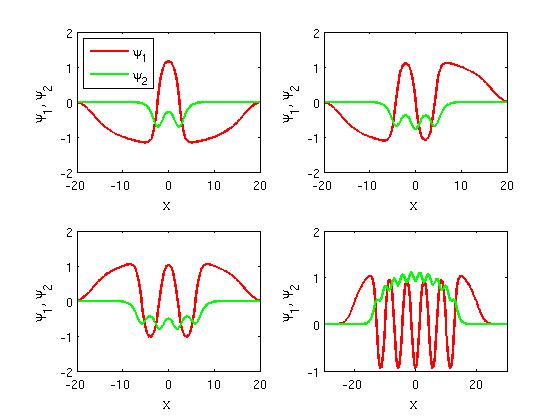}
\caption{(Color Online) A few in phase DBS states. For the two (top left), 
three (top right) and four (bottom left) dark-bright soliton states, $\mu_1=1.5,\ \mu_2=1$. For the ten dark-bright soliton states (bottom right), $\mu_1=2.25,\ \mu_2=1.5$.}
\label{dbsif}
\end{center}
\end{figure}

\begin{figure}
\begin{center}
\includegraphics[scale=0.46]{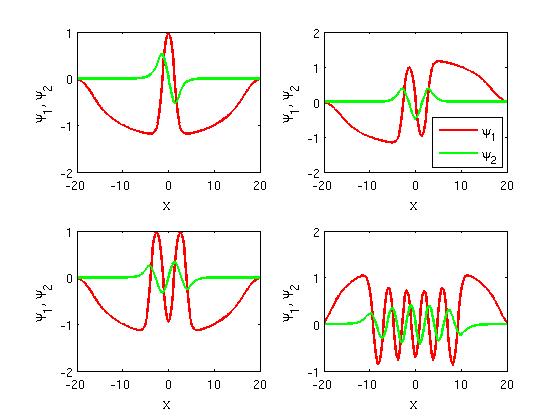}
\caption{(Color Online) A few out of phase DBS states in a similar format as in the previous
figure. For the two, three and four dark-bright soliton states, $\mu_1=1.5,\ \mu_2=1$. For the ten dark-bright soliton states, $\mu_1=1.8,\ \mu_2=1.5$.}
\label{dbsoff}
\end{center}
\end{figure}

We trace stationary states of DBSs using Newton's method, applied
to the corresponding steady state problem of Equations~(\ref{BEC}).
A key to the convergence in this regard is a suitable initialization
of the fixed point algorithm. There are two chemical potentials in the equation.
The idea of continuation from the linear limit 
is to couple a state $|n \rangle$ and $|0 \rangle$ for the in-phase DBS from 
the linear limit of quantum harmonic oscillator. For out-of-phase DBS states, 
we can couple the linear states $|n \rangle$ and $|n-1 \rangle$. 
With the starting chemical potentials  suitably chosen slightly 
above the linear limit, a continuation of the relevant states 
in the chemical potentials can be
performed. 
A few examples of both in-phase and out-of-phase DBS stationary states are 
shown in Figure~\ref{dbsif} and ~\ref{dbsoff} respectively.
The few DBS states are in line with the states reported
in~\cite{peter2}, while a discussion of DBS consisting
of many waves
and possible analytical
DBS-lattice solutions based on elliptic functions are given in~\cite{tsitoura}.

%\bibliography{references}

\subsection{Order parameter and simulation methods}

From the stationary multi-DBS state, we can clearly discern 
that the amplitudes of the bright solitons at the edges
are somewhat smaller than those around the center. This renders 
the measurement of the bright soliton locations more challenging, 
especially during the time evolution process. In this work, we only measure the locations of the dark solitons similarly to what we did in Section~\ref{peak} 
but now with a somewhat smaller cut-off of $\Delta|\psi|^2>0.5$. Then, 
we can extend the order parameter of single component dark solitons to 
the case of dark-bright solitons too. However, to confirm the genuine
two-component nature of the observed dynamics,
we have checked carefully that bright solitons are indeed following their 
dark soliton siblings in our simulations. We will discuss this 
further in section~\ref{resultdbs}. To be able to more clearly identify 
bright solitons, we chose to simulate the out-of-phase dark-bright soliton 
system. For the PDE simulation, each bright soliton was initialized with the best fit amplitudes. For the ODE and MC simulations, things are a bit more 
complicated since the interaction potentials of Equations~(\ref{UDD})-(\ref{UDB}) apply only to equal amplitude DBS pairs. Therefore, we have made an approximation 
by using the average of the best fit amplitudes of all bright solitons for each bright soliton in the ODE and PAMC simulations. This naturally 
introduces some error, but we have systematically 
checked that this doesn't substantially affect our results. For example, 
we computed the lattice constants using ODE and PAMC and found that they agree reasonably well with the PDE lattice constant. The PDE lattice constant is 2.07 while the ODE lattice constant is 2.10 and the PAMC lattice constant is found to be 2.13. We used each method's lattice constant for the respective 
simulations to compute the order parameter in a self-consistent
fashion. Moreover, we have also carefully checked that for the same disorder realization, our ODE simulation can well capture the PDE dynamics. 
This will also be discussed in section~\ref{resultdbs}.

Finally, we briefly summarize our simulation parameters for the DBS simulation. For the PDE simulation, we used 
$\mu_1=1.8,\ \mu_2=1.5,\ N=10,\ \omega=0.1,\ dx=0.05,\ dt=0.002,\ t=200.$
The reason for using different trapping frequencies for the two 
separate settings of dark and dark-bright solitary waves is 
because we are following the parameters of the original works
focusing, respectively, on them in~\cite{DS1,peter2}. In this way, 
we can benchmark some of our results against the original papers
whenever possible, e.g., as concerns 
the all-in-phase soliton oscillation frequencies, stationary states, 
equilibrium positions etc. In any event, as mentioned previously
this is simply a matter of scaling of 
length scales and should not affect our main results.
For each sample, we record $e_k$ and measure the state and the order parameter $m$ over each time period 1. Again, our ODE parameters are the same except 
for $dt=0.01$ and we also record our states over each time period 0.01  
for the ODE simulation. In our PAMC simulation, we used $R_0=5\times10^4,\ N_T=101$ and $N_s=10$. We checked that our simulation again passed the equilibrium criteria of \cite{TBC}. Having presented the general framework, we now
turn to a systematic reporting of the relevant results.

\section{Results}
\label{result}
\subsection{One component dark solitons}
\label{resultds}
\begin{figure}
\begin{center}
\includegraphics[scale=0.46]{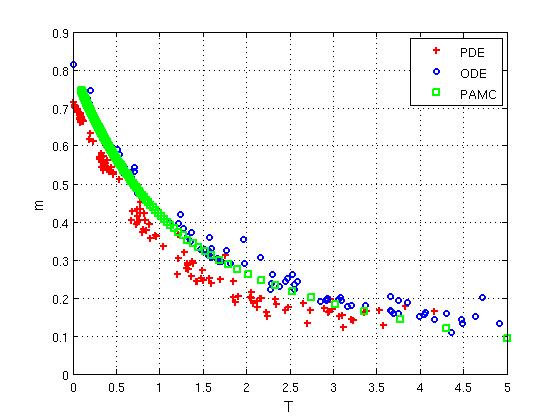}
\caption{(Color Online) The monotonically decreasing behavior of
the order parameter as a function of our kinetically
defined $T$ for the dark soliton system.}
\label{DSOP}
\end{center}
\end{figure}

The main scope of our study concerns the examination 
of how the order parameter changes as a function of our kinetically defined temperature. 
Figure~\ref{DSOP} shows how the order parameter changes with $T$. {Here we have generalized our notation of $T$ to stand for temperature for the 
MC method and for average initial kinetic energy per particle for the ODE and PDE simulations. Since the two quantities should have the same scale, this justifies the use of the same notion for simplicity. We will refer to 
this quantity as ``kinetic temperature''.} 
We can see that the order parameter features a monotonic decay as
$T$ grows. It is interesting to see that the three different methods 
(PDE, ODE and PAMC) agree reasonably well with each other in predicting
this fundamental trait. There is a gradual transition between the ordered phase and the disordered phase with an energy scale of about 1 (in our 
dimensionless units). This suggests
the existence of a modification of the system's behavior
from a highly ordered one (near unity values of $m$) to a quite disordered
one (values of $m$ around or below $0.1$). This transition
seems to be smooth and gradual and does not feature the characteristics
of standard thermodynamic phase transitions. This indeed may be 
reasonable to expect in our 1d system, although whether such genuine 
transitions may exist for different solitonic multi-particle states e.g. 
in higher dimensions remains a question worth exploring.

\begin{figure}
\begin{center}
\includegraphics[scale=0.46]{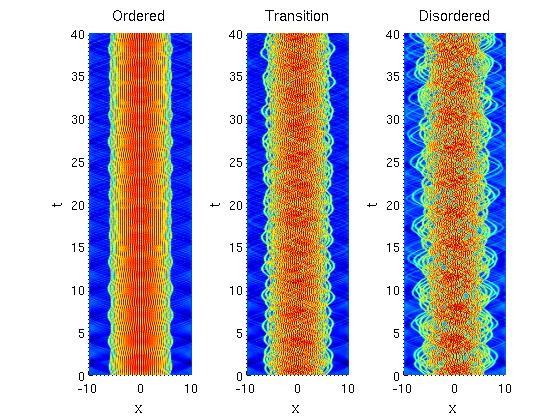}
\caption{(Color Online) Typical dynamics of the dark solitons in the ordered, transition 
(intermediate) and disordered regimes as captured by case
examples of the PDE simulation.}
\label{DSPDE}
\end{center}
\end{figure}

\begin{figure}
\begin{center}
\includegraphics[scale=0.46]{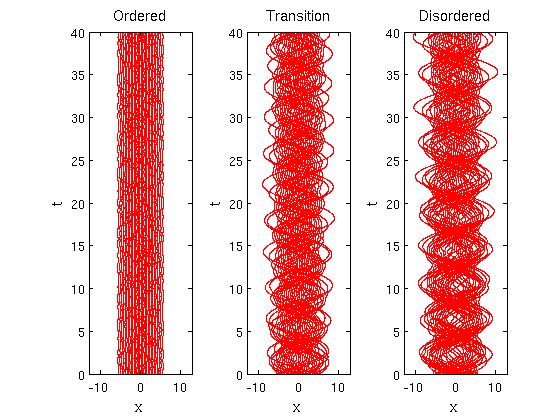}
\caption{(Color Online) Typical dynamics of the dark solitons in the ordered, transition 
(intermediate) and disorder regimes, as illustrated within the ODE simulation.}
\label{DSODE}
\end{center}
\end{figure}

The three different regimes of the dark solitons are clearly 
discernible in our simulations: there exists the ordered phase, the transition 
(intermediate) phase and the disordered phase. So it is interesting to know what the typical dynamics look like in each of these three different regimes. Figure~\ref{DSPDE} shows three typical time evolutions of the dark solitons in these
respective regimes from the PDE simulation. A similar result but from the ODE simulation is shown in Figure~\ref{DSODE}. It is clear from the PDE figure that in the ordered regime, the dark solitons don't cross each other. In the transition regime, they start to cross each other once in a while. In the disordered regime, they do crossing frequently.
In this
case, the highly energetic (both in the PDE and in the ODE) soliton particles
resemble those of a  ``gas''.

\subsection{Dark-bright solitons}
\label{resultdbs}
\begin{figure}
\begin{center}
\includegraphics[scale=0.46]{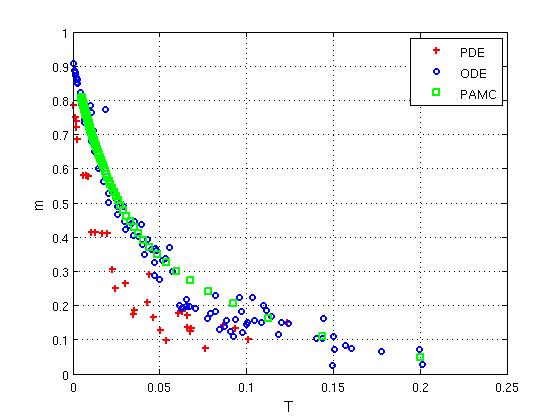}
\caption{(Color Online) The monotonically decreasing dependence of the 
order parameter as a function of our kinetically defined
$T$ for the DBS system.}
\label{DBSOP}
\end{center}
\end{figure}

\begin{figure}
\begin{center}
\includegraphics[scale=0.46]{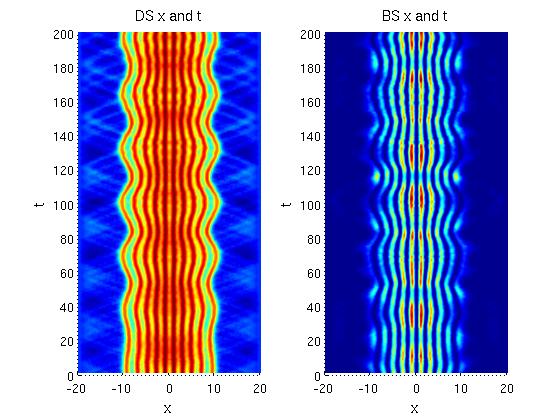}
\caption{(Color Online) Typical dynamics of the ordered state for the DBS system, 
resulting from the PDE simulation. The space-time evolution of the field
is shown in the two components.}
\label{PDEorder}
\end{center}
\end{figure}

\begin{figure}
\begin{center}
\includegraphics[scale=0.46]{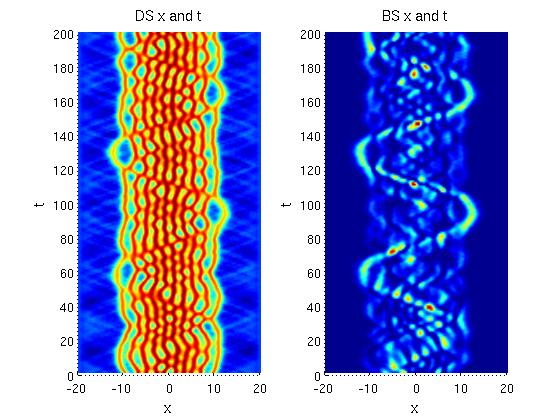}
\caption{(Color Online) Typical dynamics of the transition (intermediate)
state for the DBS system, 
from the PDE simulation.}
\label{PDEtransition}
\end{center}
\end{figure}

\begin{figure}
\begin{center}
\includegraphics[scale=0.46]{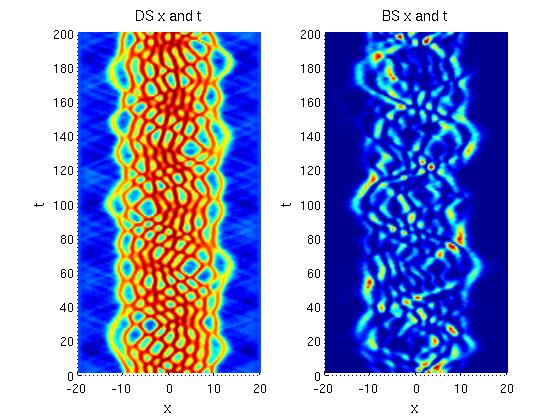}
\caption{(Color Online) Typical dynamics of the disordered state for the DBS system, 
again showing both components space-time evolution from the PDE simulation.}
\label{PDEdisorder}
\end{center}
\end{figure}

\begin{figure}
\begin{center}
\includegraphics[scale=0.46]{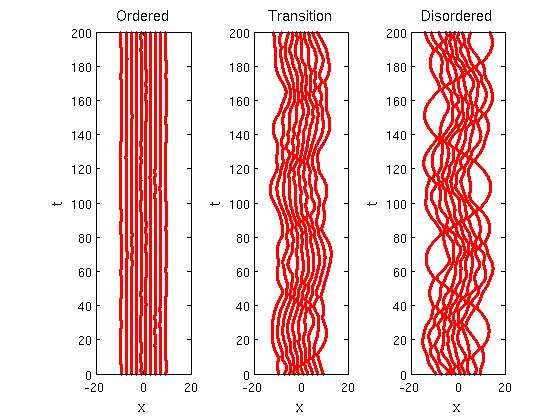}
\caption{(Color Online) Typical dynamics of the ordered, transition and disordered states for the DBS system, as resulting from the ODE simulation.}
\label{ODEotd}
\end{center}
\end{figure}

Here we show similar results of how the order parameter changes as
our kinetic temperature increases for the DBSs. Figure~\ref{DBSOP} shows how the order parameter changes with $T$.
There is again a gradual transition between the ordered phase and the disordered phase with an energy scale of about 0.05 for the parameters selected here. Once again, the overall 
trend of the PDE and ODE, as well as the PAMC is fairly similar. 
Nevertheless, the PDE seems to be somewhat less ordered than the other
two, conceivably because of the enhanced role of the additional
degrees of freedom and the more complex nature of the associated
dynamics in this two-component setting. 
 Some typical PDE and ODE ordered, transition (i.e., intermediate)
and disordered states are shown in Figures~\ref{PDEorder},
\ref{PDEtransition}, \ref{PDEdisorder} and \ref{ODEotd}.

\begin{figure}
\begin{center}
\includegraphics[scale=0.46]{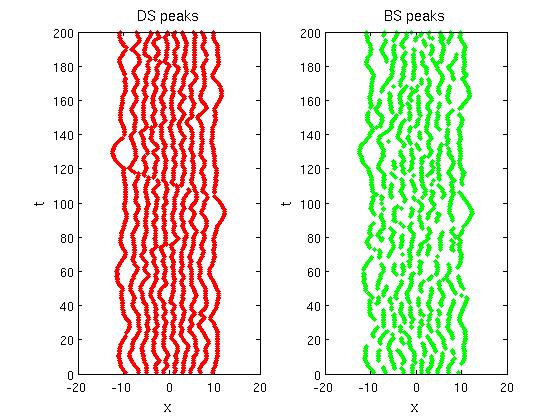}
\caption{(Color Online) 
The same states as \ref{PDEtransition}, but instead of showing 
density, we plotted the peaks of the states from the dark (left) and
bright (right) components.}
\label{PDE2}
\end{center}
\end{figure}

We also want to point out here that Figure~\ref{PDEtransition} and Figure~\ref{PDEdisorder} are worth commenting upon further in that some of the bright solitons seem to be less visible due to the highly collisional
nature of the dynamics. We therefore checked the dynamical 
stability of the stationary dark-bright soliton state. In accordance
with the results of~\cite{tsitoura}, we found there is some instability but nevertheless the corresponding growth rate is rather small, i.e., small enough that
over the time scales reported herein, the resulting weak dynamical instabilities
(of stationary states) have not yet set in. Our detailed examination of this feature suggests
that during the intermediate, as well as disordered phase dynamics, 
the collisional dynamics may develop high amplitudes, thus rendering 
some of the bright solitons less visible in the space-time plots of 
e.g.~\ref{PDEtransition} and \ref{PDEdisorder}. To shed further light on this
issue, 
we have looked at the peaks of the dark (after being subtracted
from the ground state background)
and bright soliton components. 
The result of the same states as Figure~\ref{PDEtransition} is shown in Figure~\ref{PDE2}, which clearly attests to the fact that the bright solitons are 
indeed following suit with respect to their dark soliton partners. 
Similar results are obtained for states in other disorder realizations.
Nevertheless, this phenomenology of amplitude enhancement and 
apparent ``mass exchange'' may be worth exploring further and may
be, in part, related to (a generalization of) 
the two DBS self-trapping phenomenology recently reported
in~\cite{karamatskos}.

\begin{figure}
\begin{center}
\includegraphics[scale=0.46]{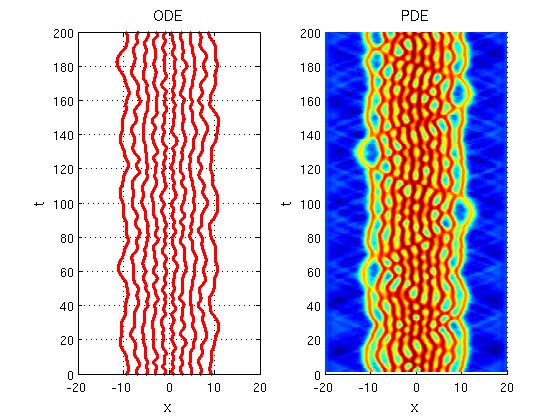}
\caption{(Color Online) The same states as \ref{PDEtransition}, but the left panel is now the ODE simulation with the same disorder realization as the corresponding
PDE simulation. The dark soliton component is shown on the right panel. 
The two simulations have the same initial kinetic energy. Note that 
the ODE can accurately capture the PDE dynamics.}
\label{PDEODE2}
\end{center}
\end{figure}

Finally, it is interesting to check whether the ODE simulation can match the peaks in the PDE simulation for the same disorder realization. A typical result of the ODE and PDE simulation of the same disorder realization as Figure~\ref{PDEtransition} with the same initial kinetic energy is shown in Figure~\ref{PDEODE2}. We have shown the trajectories of the dark-bright solitons of the ODE simulation on the left panel and the dark soliton component of the PDE simulation on the right panel. It is clear from the figure that even though we have made some approximations for the particle interactions, the ODE simulation nevertheless 
captures fairly accurately 
the dynamics of the full PDE simulation. Similar results were found for 
dark-bright solitons in other regimes and also for the simpler case
involving solely dark solitons.

\section{Conclusions and Future Challenges}
\label{conclusion}

In the present work, we did a systematic numerical simulation of the 
states of the one-dimensional dark soliton and dark-bright soliton lattices 
in the large density limit using PDE, ODE and PAMC simulations. 
We identified regimes where the dynamics, as characterized both
by a concrete, yet empirical,
 order parameter and by a direct inspection of the 
space-time evolution appears ordered, as well as ones where it appears
highly disoredered and also monitored intermediate (``transitional'')
regimes between these two.
The three methods of our numerical choice gave similar results, verifying
the good agreement between our different dynamical descriptions. For our
defined order parameter, we found that it continuously decreases when 
our kinetically defined temperature parameter increases. Nevertheless,
in our current formulation of the problem and although the different
end states for low and high kinetic energy can be considered as
``soliton crystals'' and ``soliton gases'', respectively, no genuine
phase transitions were identified in our one-dimensional, one-
and two-component formulations.

Our analysis poses a number of interesting questions for future study.
One such concerns whether a more rigorous (or numerically assisted)
characterization of the thermodynamic properties can be provided for
our effective wave-particle system via e.g. the transfer integral
method~\cite{ti1,ti2}. For instance, in the dark soliton case, the
effective particle system is a perturbed form of a Toda lattice,
while for the classical (integrable) Toda lattice, transfer integral
based techniques have been used to compute the partition function
and thermodynamic quantities, e.g., in~\cite{ti3,ti4}. The use of such
techniques even in a numerical form could provide a definitive 
answer in the question as to whether phase transitions may or
may not exist in the present setting. Additionally, it would
be particularly interesting to generalize relevant considerations
to higher dimensional settings. In particular, 
a similar effective description can be formulated in the case of
a gas of trapped vortices in quasi-2d BECs,
where  a reduced particle description in the large density
limit is also available; see e.g.~\cite{DS2d}. Thus, once again, the use of 
suitable
order parameters can be used to identify the thermodynamics
properties of the relevant vortex cloud. These questions are
currently under consideration and will be reported in future
studies.

\begin{acknowledgments}

W.W. acknowledges support from Prof. Jon Machta of the Physics department of UMass Amherst via NSF (Grant No.~DMR-1208046). 
P.G.K. gratefully acknowledges the support of 
NSF-DMS-1312856, as well as from
the US-AFOSR under grant FA950-12-1-0332,  
and the ERC under FP7, Marie
Curie Actions, People, International Research Staff
Exchange Scheme (IRSES-605096).
P.G.K.'s work at Los Alamos is supported in part by the U.S. Department
of Energy. We thank Jon Machta for helpful discussions, especially 
regarding the Monte Carlo simulations, Dimitri Frantzeskakis
for numerous fruitful discussions on the themes of dark and dark-bright
solitons and also Evangelos Karamatskos for discussions on the
dynamics of multiple dark-bright solitons.

\end{acknowledgments}

% use  \cite{Aa10} for references;

%\bibliographystyle{plain}
%\bibliography{references}

\end{document}